\documentclass[twocolumn,prl,aps,showpacs,floatfix,superscriptaddress,preprintnumbers,amsmath,amssymb]{revtex4}

\usepackage{graphicx}

\begin{document}

\newcommand{\cso}{Cu$_2$OSeO$_{3}$}
\newcommand{\mfs}{Mn$_{1-x}$Fe$_{x}$Si}
\newcommand{\mcs}{Mn$_{1-x}$Co$_{x}$Si}
\newcommand{\fcs}{Fe$_{1-x}$Co$_{x}$Si}
\newcommand{\dto}{Dy$_{2}$Ti$_{2}$O$_{7}$}

%%%%%%%%%%%%%%%%%%%%%%%%%%%%%%%%%%%
\newcommand{\ozz}{$\langle100\rangle$}
\newcommand{\ooz}{$\langle110\rangle$}
\newcommand{\ooo}{$\langle111\rangle$}
\newcommand{\too}{$\langle211\rangle$}

\newcommand{\mb}{$\mu_0\,M/B$}
\newcommand{\dmdb}{$\mu_0\,\mathrm{d}M/\mathrm{d}B$}
\newcommand{\ddmddb}{$\mathrm{\mu_0\Delta}M/\mathrm{\Delta}B$}
\newcommand{\cm}{$\chi_{\rm M}$}
\newcommand{\cac}{$\chi_{\rm ac}$}
\newcommand{\rechi}{${\rm Re}\,\chi_{\rm ac}$}
\newcommand{\imchi}{${\rm Im}\,\chi_{\rm ac}$}

\title{Long wavelength helimagnetic order and skyrmion lattice phase in {\cso}}

\author{T. Adams}
\affiliation{Technische Universit\"at M\"unchen, Physik-Department E21, D-85748 Garching, Germany}

\author{A. Chacon}
\affiliation{Technische Universit\"at M\"unchen, Physik-Department E21, D-85748 Garching, Germany}
\affiliation{Forschungsneutronenquelle Heinz Maier Leibnitz (FRM II),  Lichtenbergstr. 1, 85748 Garching, Germany}

\author{M. Wagner}
\affiliation{Technische Universit\"at M\"unchen, Physik-Department E21, D-85748 Garching, Germany}

\author{A. Bauer}
\affiliation{Technische Universit\"at M\"unchen, Physik-Department E21, D-85748 Garching, Germany}

\author{G. Brandl}
\affiliation{Technische Universit\"at M\"unchen, Physik-Department E21, D-85748 Garching, Germany}
\affiliation{Forschungsneutronenquelle Heinz Maier Leibnitz (FRM II),  Lichtenbergstr. 1, 85748 Garching, Germany}

\author{B. Pedersen}
\affiliation{Forschungsneutronenquelle Heinz Maier Leibnitz (FRM II),  Lichtenbergstr. 1, 85748 Garching, Germany}

%\author{R. Georgii}
%\affiliation{Technische Universit\"at M\"unchen, Physik-Department E21, D-85748 Garching, Germany}
%\affiliation{Forschungsneutronenquelle Heinz Maier Leibnitz (FRM II),  Lichtenbergstr. 1, 85748 Garching, Germany}

\author{H. Berger}
\affiliation{Ecole Polytechnique Federale Lausanne, CH-1015 Lausanne, Switzerland}

\author{P. Lemmens}
\affiliation{Institute for Condensed Matter Physics, TU Braunschweig, D-38106 Braunschweig, Germany}
 
\author{C. Pfleiderer}
\email{christian.pfleiderer@frm2.tum.de}
\affiliation{Technische Universit\"at M\"unchen, Physik-Department E21, D-85748 Garching, Germany}

\date{\today}

\begin{abstract}
We report a long-wavelength helimagnetic superstructure in bulk samples of the ferrimagnetic insulator {\cso}. The magnetic phase diagram associated with the helimagnetic modulation inferred from small angle neutron scattering and magnetisation measurements includes a skyrmion lattice phase and is strongly reminiscent of MnSi, FeGe and {\fcs}, i.e., binary isostructural siblings of {\cso} that order helimagnetically. The temperature dependence of the specific heat of {\cso} is characteristic of nearly critical spin fluctuations at the helimagnetic transition. This provides putative evidence for effective spin currents as the origin of enhancements of the magneto-dielectric response instead of atomic displacements considered so far.
\end{abstract}


\vskip2pc

\maketitle
%%%%%%%%%%%%%%%%%%%%%%%%%%%%%%%%

Major efforts have been made in recent years to unravel the nature of the magneto-electric coupling in multiferroic materials \cite{Erenstein:Nature2006,Fiebig:JPD2005,Spaldin:Science2005,Cheong:NatMat2007}. Amongst a wide range of theoretical scenarios two mechanisms are considered most prominent. First, a coupling mediated by effective spin currents in spin spiral magnets and, second, an exchange striction mechanism in which the magneto-elastic coupling proceeds via atomic displacements. An important property believed to provide unambiguous evidence of the latter mechanism is an enhancement of the magneto-dielectric response (MDR), describing changes in the dielectric polarization in applied magnetic fields or in the presence of long range magnetic order. However, the discovery of an enhanced MDR near the magnetic transition of the insulator {\cso} appears to question this view \cite{Bos:PRB2008}. Detailed studies of the crystal structure and lattice dynamics strongly suggest the absence of spontaneous lattices strains \cite{Bos:PRB2008,Gnezdilov:LTP2010,Kobets:LTP2010,Miller:PRB2010}. 
Being a lone pair containing piezoelectric ferrimagnet this was taken as evidence of a new magneto-electric coupling mechanism. Yet, enhancements of the MDR without lattice strains are not specific to multiferroics and may represent a more general scientific challenge. For instance, the spin ice system {\dto} displays an enhanced MDR but is neither magnetically ordered nor multiferroic \cite{Saito:PRB2005}.

{\cso} is ideally suited to search for the origin of enhancements of the MDR without lattice strains. It crystallises in the non-centrosymmetric B20 structure, space group P2$_1$3 \cite{Meunier:JAC1976}, which structurally allows ferroelectricity. The unit cell is composed of three building blocks \cite{Effenberger:MC1986}. The first and the second building block are given by a square pyramidal and a trigonal bipyramidal CuO$_5$ unit in a three to one ratio, respectively. The third building block is a lone pair containing a tetrahedral SeO$_3$. Magnetisation measurements and powder neutron diffraction have established ferrimagnetic order of the Cu$^{2+}$ moments below $T_c=58.8\,{\rm K}$, where three ferromagnetically aligned Cu moments pair up antiferromagnetically with a fourth Cu moment  \cite{Kohn:JPSJ1977,Bos:PRB2008}. The exchange coupling is given by $J_{\rm FM}=-50\,{\rm K}$ and $J_{\rm AFM}=68\,{\rm K}$ for the ferromagnetic and antiferromagnetic exchange, respectively, consistent with the Kanamori-Goodenough rules \cite{Belesi:PRB2010}. 

Yet, the description in terms of ferrimagnetic order seems incomplete. First, well below $T_c$ the magnetisation increases almost linearly with increasing field before saturating above $\sim120\,{\rm mT}$ without evidence for a spontaneous uniform magnetic moment at $B=0$ \cite{Bos:PRB2008,Huang:PRB2011,Belesi:JPCS2011}. If the behaviour for $B\lesssim120\,{\rm mT}$ would be due to magnetic domains the slope and the onset of saturation would sensitively reflect demagnetising fields and depend on the sample shape which is not observed. Second, small changes of slope in the initial increase of $M(B)$ suggest that the magnetic state is more complex \cite{Huang:PRB2011}. Third, the ferrimagnetic order is incompatible with the crystal structure of {\cso} and a symmetry lowering transition would be expected which is not observed \cite{Bos:PRB2008,Belesi:PRB2010,Huang:PRB2011,Miller:PRB2010}.
Finally, a detailed magnetic phase diagram of bulk samples was recently determined in magnetisation and electric polarisation measurements for a {\ooo} axis \cite{Tokura:2011,Seki:2012}. Based on Lorentz force microscopy in thin {\cso} samples helimagnetic order and a skyrmion lattice phase were attributed, but the phase diagram of the thin samples had completely different phase boundaries. Moreover, for the thin samples the helical order was found to propagate along {\ooz}, which is not supported by the B20 symmetry in bulk samples contrasting a key assumption in Ref.\,\cite{Seki:2012,Seki:2012-full}.

In this Letter we report a helimagnetic superstructure in \textit{bulk samples} of {\cso}, which resolves all of the questions listed above and suggests strongly that the enhanced MDR arises from spin currents due to nearly critical helimagnetic spin fluctuations. The helimagnetic order in {\cso} relates thereby to binary transition metal (TM) compounds such as MnSi and FeGe, in which a hierarchy of three energy scales arises in the B20 crystal structure \cite{Landau}. On the strongest and second strongest scale these are ferromagnetic exchange and Dzyaloshinsky-Moriya interactions, respectively, generating a long-wavelength helimagnetic modulation. The propagation direction of the helix is thereby the result of very weak magnetic anisotropies due to higher order spin-orbit coupling terms on the weakest scale. Most remarkable perhaps, a skyrmion lattice phase was recently discovered in the magnetic phase diagram of the binary TM B20 compounds \cite{mueh09,Muenzer:PRB2010,yu:Nature2010,yu:NatureMaterials2011}, which gives rise to a novel emergent electrodynamics \cite{joni10,Schulz:NaturePhysics2012}. 

For our study a large {\cso} single crystal ($\sim0.2\,{\rm g}$) was grown by chemical vapour transport. Further properties of samples from the same growth were reported in Refs.\,\cite{Gnezdilov:LTP2010,Kobets:LTP2010,Belesi:PRB2010,Huang:PRB2011,Maisuradze:PRB2011} which also describe technical details of the growth procedure.  An excellent sample quality was established by means of neutron diffraction at RESI at FRM\,II, notably an absence of impurity phases well below 1\,\% and a lattice mosaic spread smaller than the resolution limit of $0.02^{\circ}$. Using  a neutron wavelength $\lambda_n=1.0408\,{\rm \AA}$ a simultaneous refinement of 60 Bragg spots provided lattice parameters $a=b=c=(8.9199\pm0.00104)\,{\rm \AA}$ and $\alpha=\beta=\gamma=90^{\circ}$, in excellent agreement with the literature \cite{Meunier:JAC1976,Effenberger:MC1986,Bos:PRB2008}. 

\begin{figure}
\includegraphics[width=0.45\textwidth]{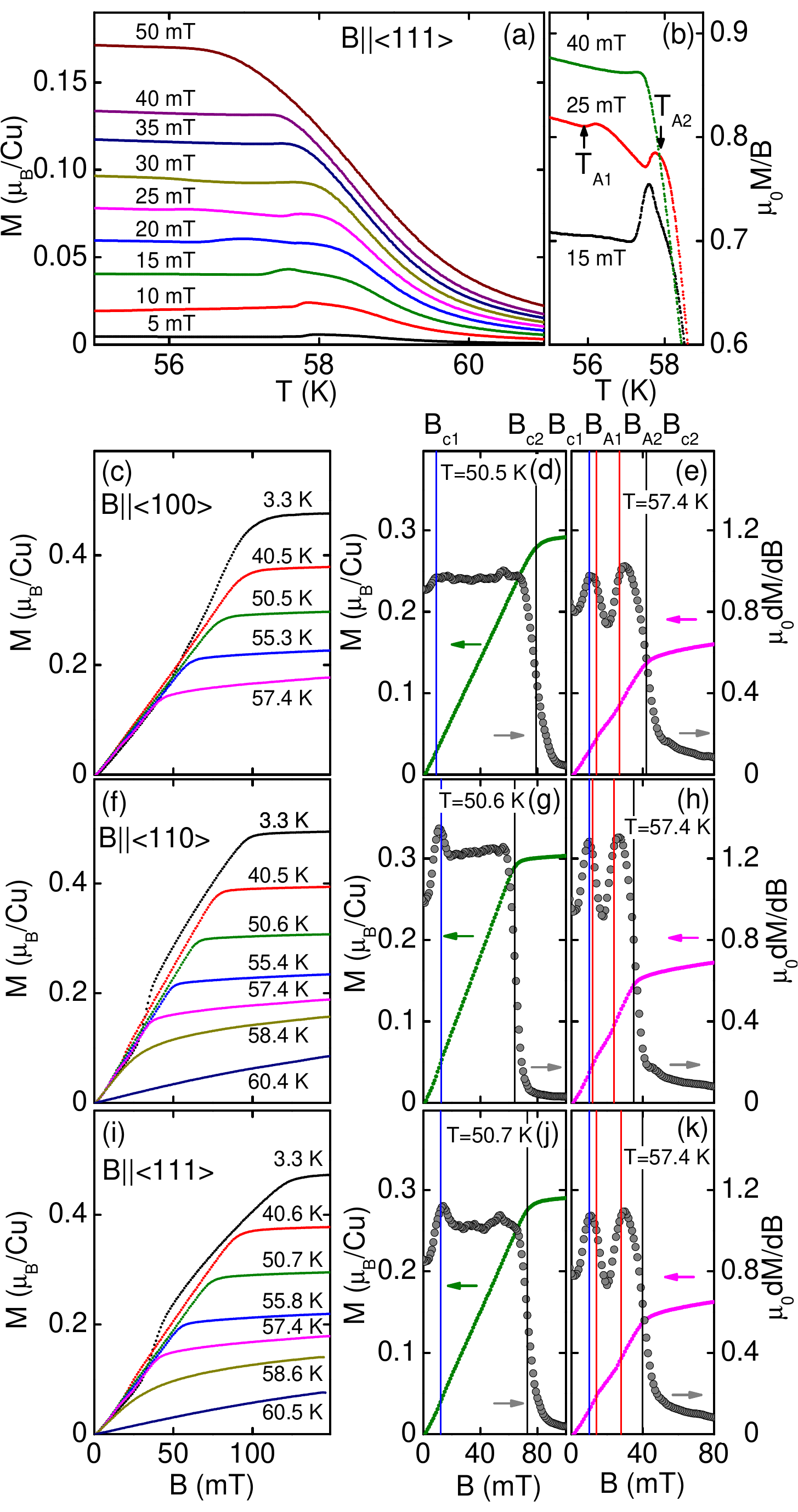}
\caption{(Colour online) Magnetisation of single crystal {\cso} for various crystallographic directions. (a) Temperature dependence of the magnetisation in the vicinity of $T_c$. (b) Ratio $\mu_0\,M/B$ versus temperature revealing the features characteristic of the transition to the A-phase. Panels (c) through (j): Magnetisation as a function of field at various temperatures. Panels on the right hand side show typical data just below $T_c$, where a clear minimum in {\dmdb}, calculated from the magnetisation, is observed in the A-phase.
}
\label{Fig1}
\end{figure}

The small angle neutron scattering (SANS) measurements were performed at the diffractometer MIRA2 at FRM\,II using a neutron wavelength $\lambda_n=5\,{\rm \AA}\pm5\%$. The neutron beam was collimated with an aperture, $3\times3\,{\rm mm^2}$, placed $\sim$1.5\,m in front of the sample and a second aperture, $0.6\times0.8\,{\rm mm^2}$, placed 0.38\,m in front of the sample. The scattered neutrons were recorded with a CASCADE detector \cite{Klein:NIMA2011} located 2\,m behind the sample. For our measurements the sample was cooled with a Sumitomo pulse tube cooler, where a bespoke sample stick permitted automated rotation with respect to an axis vertical to the incident neutron beam. The rotation axis coincided with a crystallographic {\ooz} axis. The magnetisation was measured down to 3\,K with an Oxford Instruments vibrating sample magnetometer at magnetic fields up to 9\,T. The specific heat was measured in a Quantum Design PPMS using a quasi-adiabatic long-pulse technique. Laue x-ray diffraction was used to orient the sample.

Shown in Fig.\ref{Fig1}\,(a) are typical magnetisation data as a function of temperature in the vicinity of $T_c$.  Well above $T_c$ a strong Curie-Weiss dependence with a large fluctuating moment, $\mu_{\rm CW}\approx1.5\,\mu_{\rm B}/{\rm Cu}$ is observed in perfect agreement with the literature. With increasing applied magnetic field the magnetisation increases. In the vicinity of $T_c$ faint maxima develop as illustrated in Fig.\ref{Fig1}\,(b), where $M/B$ is shown for clarity. These features are analogous to MnSi \cite{Bauer:2012}, where they arise from the skyrmion lattice phase. The temperature dependence is consistent with the magnetic field dependence shown in Fig.\,\ref{Fig1}\,(c) through (k) for field along {\ozz}, {\ooz} and {\ooo}. With decreasing temperature $M(B)$ increases before reaching a saturated moment $m_s=0.48\,\mu_{\rm B}/{\rm Cu}$ at large fields. Additional changes of the slope of $M(B)$ for temperatures just below $T_c$ are best revealed in {\dmdb}, where a distinct minimum is observed in a small $T$ interval as illustrated in Fig.\,\ref{Fig1}\,(e), (h) and (k). We thereby define transition fields $B_{c1}$, $B_{A1}$, $B_{A2}$ and $B_{c2}$ as in the binary B20 compounds as shown in Fig.\,\ref{Fig1} (cf \cite{Bauer:PRB2010}).

\begin{figure}
\includegraphics[width=0.45\textwidth]{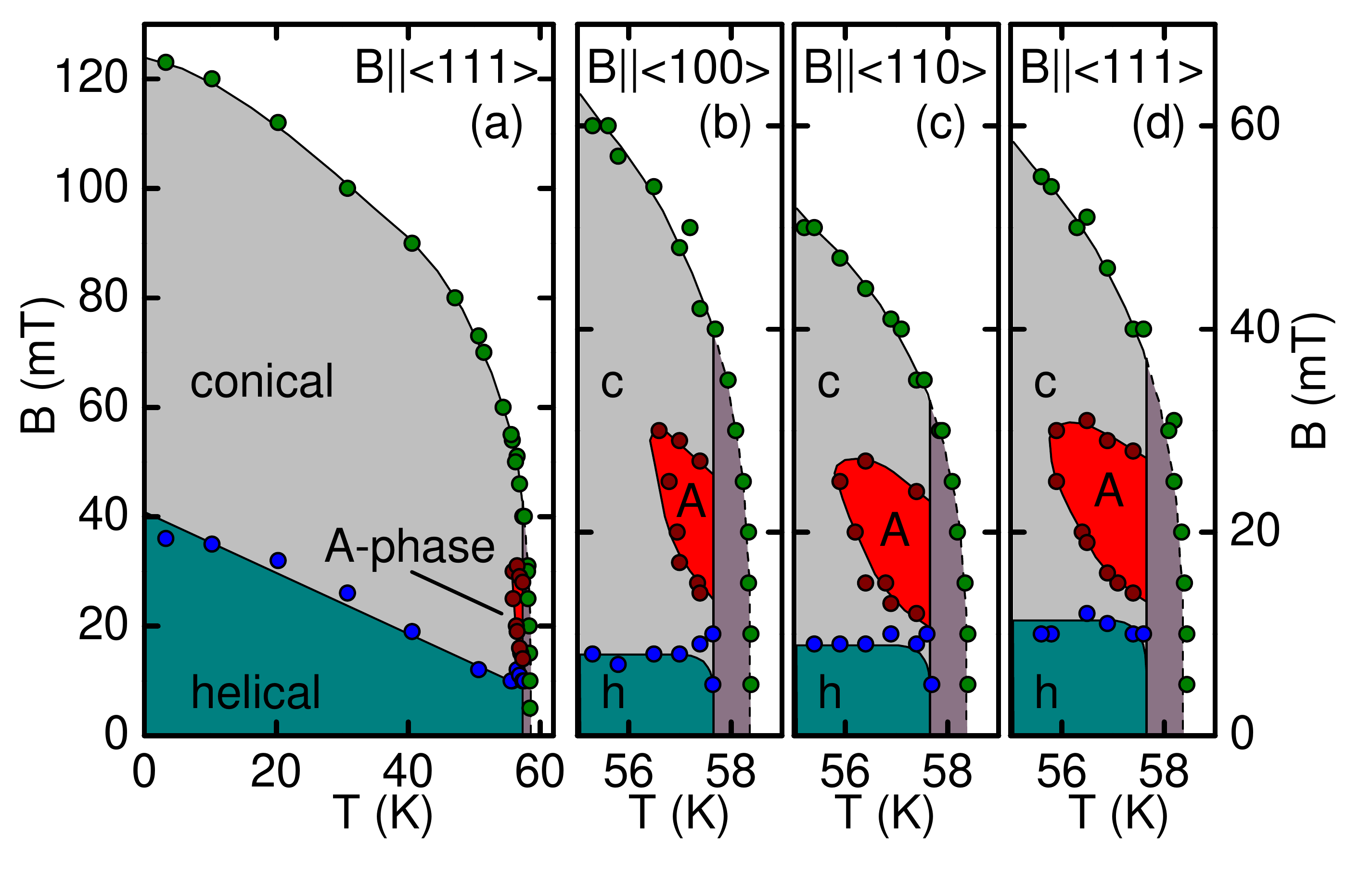}
\caption{(Colour online) Magnetic phase diagram of {\cso} as a function of applied magnetic field for various orientations inferred from the magnetisation. (a) Overview for field parallel {\ooo}. Panels (b) through (d): Phase diagram in the vicinity of $T_c$ for various orientations. Differences as a function of field are mostly due to demagnetising effects; the brown shading indicates the regime of nearly critical spin fluctuations.
}
\label{Fig2}
\end{figure}

Based on the magnetisation we obtain the magnetic phase diagram shown in Fig.\ref{Fig2}. The SANS data described below identify the following phases: (i) for $B<B_{c1}$ helimagnetic order denoted 'h', (ii) for $B_{c1}<B<B_{c2}$ conical order denoted 'c', (iii) for $B>B_{c2}$ field-polarized ferrimagnetic order, and finally (iv) a skyrmion lattice in the regime denoted 'A', just below $T_c$. We note that differences of $B_{c2}$ reflect demagnetising fields, which cannot be corrected accurately for the shape of our sample. Likewise, the field range of the skyrmion lattice phase varies weakly with field direction (Fig.\ref{Fig2}(b), (c) and (d)). However, the temperature range is clearly largest for {\ooo} and smallest for {\ozz}, consistent with the magnetic anisotropy favouring the propagation of the helical order at zero field along {\ozz} \cite{Bauer:2012,bak80,naka80}.

\begin{figure}
\includegraphics[width=0.45\textwidth]{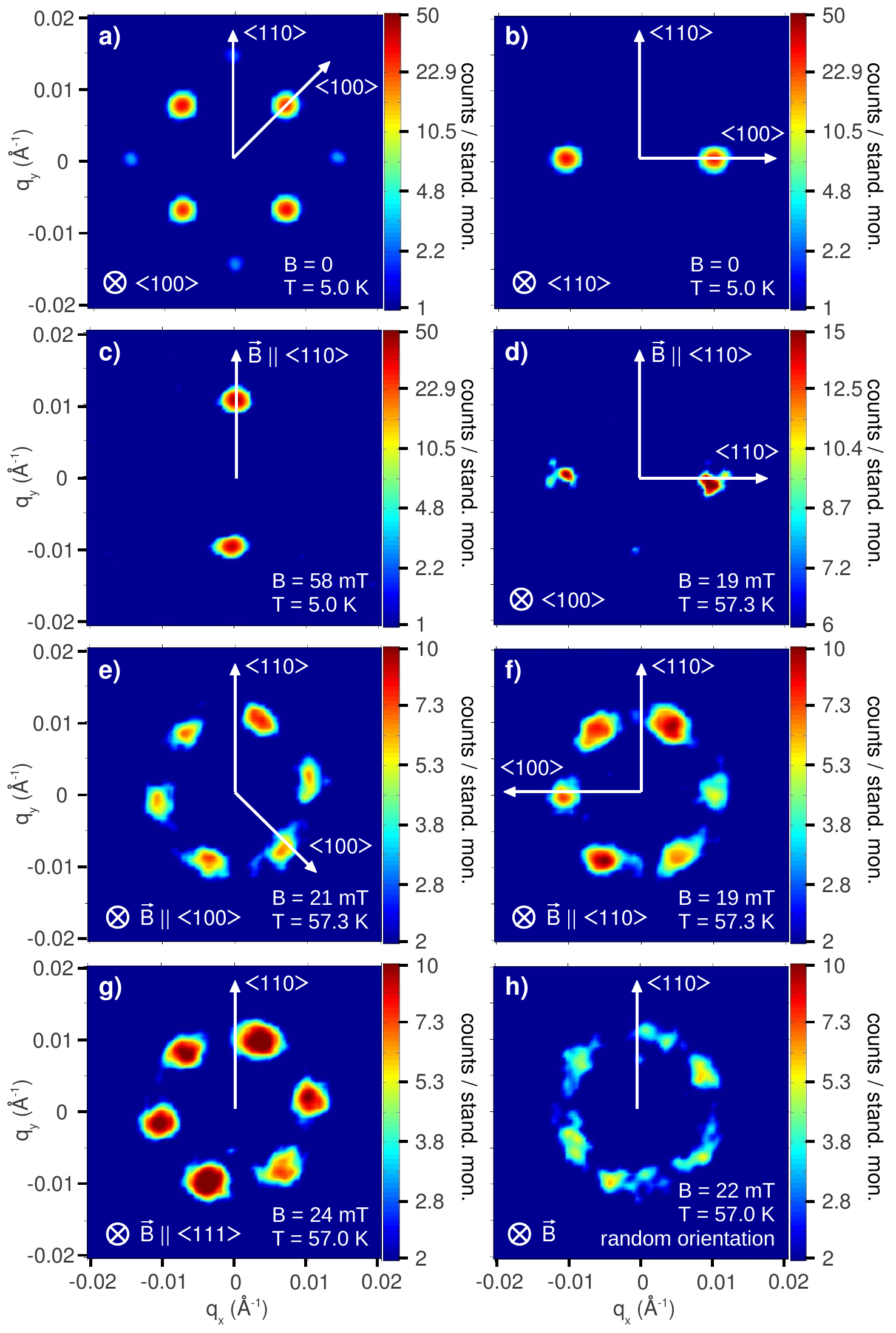}
\caption{(Colour online) Typical integrated small angle neutron scattering rocking scans in {\cso}. (a) Zero field scattering pattern along {\ozz} characteristic of helimagnetic order along {\ozz}. (b) Zero field scattering pattern along {\ooz} characteristic of a helimagnetic modulation along {\ozz}. (c) Typical scattering pattern for $B_{c1}<B<B_{c2}$ for $T\ll T_c$. (d) Scattering pattern in the A-phase for magnetic field perpendicular to the neutron beam. Panels (e) through (h): Typical scattering pattern in the A-phase for magnetic field parallel to the neutron beam for various orientations.}
\label{Fig3}
\end{figure}

Typical intensity patterns of integrated rocking scans that identify the various phases are shown in Fig.\,\ref{Fig3}. Magnetic rocking widths were small in all magnetic phases. An exception was the plane perpendicular to the applied field in the A-phase, where the precise intensity distribution was also sensitive to the field and temperature history. Future studies will have to establish whether this was the result of demagnetising fields related to the shape of our single crystal akin the shape dependence observed in MnSi single crystals \cite{Adams:PRL2011}. For $B=0$ the intensity pattern consists of well defined spots at $k\sim(0.0102\pm0.0008)\,{\rm \AA^{-1}}$ along all three {\ozz} axes, characteristic of a modulation
with a long wavelength $\lambda_{\rm h}\approx 616\pm 45\,{\rm \AA}$. This is shown in Figs.\,\ref{Fig3}\,(a) and (b), which displays the intensity pattern for neutrons parallel {\ozz} and {\ooz}, respectively. Preliminary tests with polarised neutrons suggest a homochiral helical modulation. The very weak additional spots along the {\ooz} axes in Fig.\,\ref{Fig3}\,(a) are characteristic of double scattering. By analogy with the binary B20 systems the scattering pattern at $B=0$ is  characteristic of a multi-domain single-$\vec{k}$ helimagnetic state, where spots along each {\ozz} axes correspond to different domain populations. In contrast, in MnSi the helical modulation is along {\ooo}. This is consistent with a change of sign of the leading order magnetic anisotropy in {\cso}
\cite{bak80,naka80,mueh09,Muenzer:PRB2010}, but contrasts distinctly the {\ooz} propagation direction reported for thin samples \cite{Seki:2012}.

In the range $B_{c1}<B<B_{c2}$ the zero-field pattern (Figs.\,\ref{Fig3}\,(a) and (b)) collapses into two spots parallel to the field. This is illustrated for $B=60\,{\rm mT}$ applied vertical to the neutron beam at 5\,K in Fig.\,\ref{Fig3}\,(c). Accordingly the modulation is parallel to $B$ and, in analogy with the binary B20 compounds, characteristic of a spin-flop phase also referred to as conical phase.  

\begin{figure}
\includegraphics[width=0.45\textwidth]{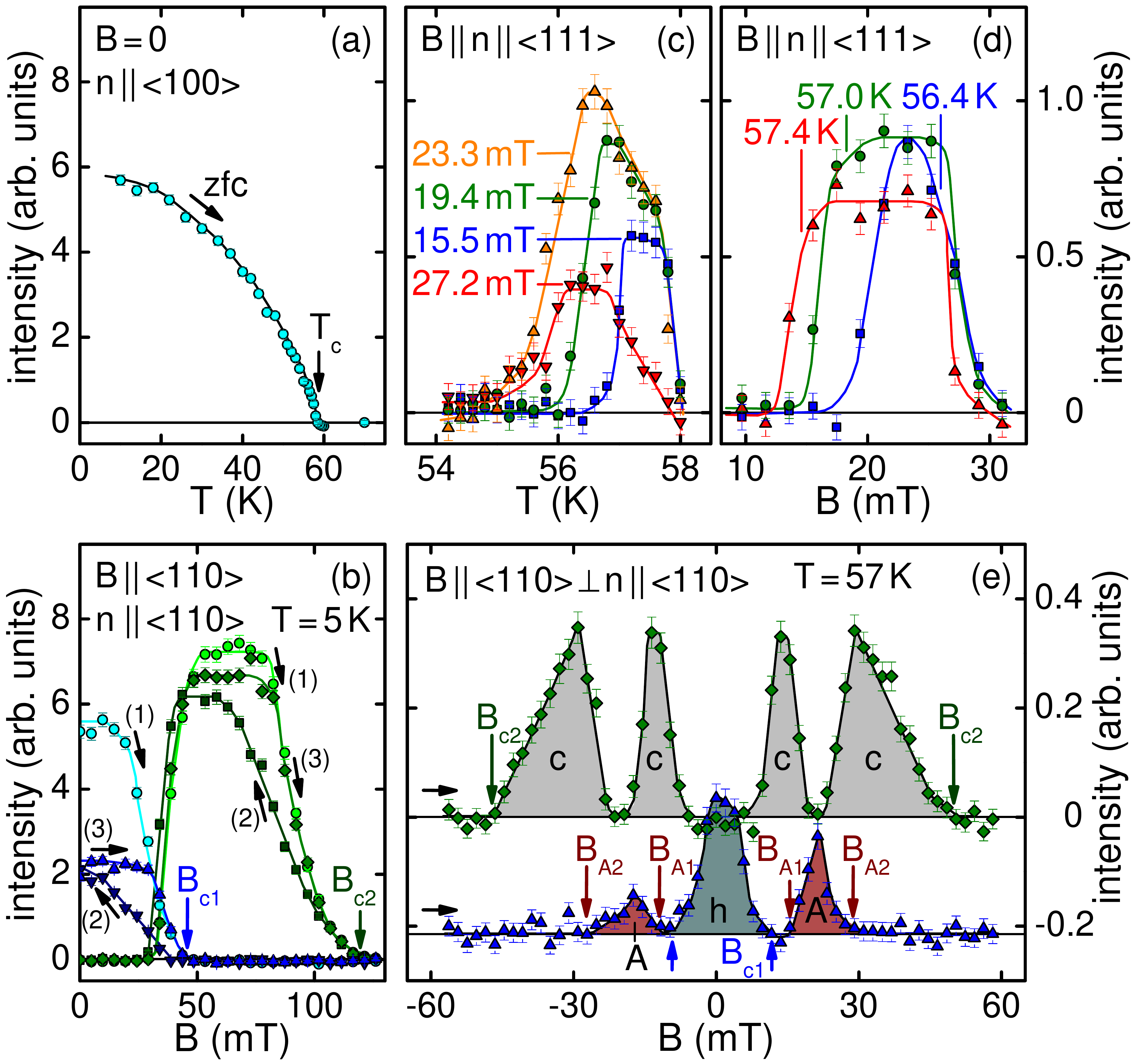}
\caption{(Colour online) Typical temperature and magnetic field dependences of peak intensities recorded in SANS (see text for details). (a) Temperature dependence of the helical order, cf. Fig.\,\ref{Fig3}\,(a). (b) Field dependence of the helical and conical state, where blue data points correspond the helical state, cf. Fig.\,\ref{Fig3}\,(b), and green data points to the conical state, cf. Fig.\,\ref{Fig3}\,(c). Labels (1), (2) and (3) mark the sequence in which sweeps were recorded; arrows show the sweep direction. Panels (c) \& (d): Temperature and field dependence in the A-phase, cf. Fig\,\ref{Fig3}\,(g). (e) Field dependence in the temperature range of the A-phase; helical order (blue, marked h), conical order (green, marked c) and A-phase (blue marked A), cf. Fig.\,\ref{Fig3}\,(b), (c) and (d).  
}
\label{Fig4}
\end{figure}

In the A-phase, finally, the intensity pattern consists essentially of a ring of six spots perpendicular to the applied magnetic field, regardless of the orientation of the single crystal with respect to the field direction. This is illustrated in Figs.\,\ref{Fig3}\,(d) through (h). We begin with panel (d) which demonstrates, that the pattern for magnetic field perpendicular to the neutron beam is also perpendicular to the field. Further, Figs.\,\ref{Fig3}\,(e) through (h) show the six-fold pattern for field parallel to the neutron beam. The orientation of the six-fold pattern in the plane perpendicular to the field is thereby roughly aligned along {\ozz}, consistent with very weak magnetic anisotropy terms that are sixth order in spin-orbit coupling and the effects of demagnetising fields (see e.g. \cite{mueh09,Muenzer:PRB2010}). 

As demonstrated for the binary B20 compounds the six-fold pattern arises from a triple-$\vec{k}$ state, with $\sum_i \vec{k}_i=0$, coupled to the uniform magnetisation and stabilised by thermal Gaussian fluctuations (a single-$\vec{k}$ modulation perpendicular to the applied field is energetically least favourable \cite{mueh09}). The topology of this peculiar multi-$\vec{k}$ state is that of a skyrmion lattice, i.e. the winding number is -1 per magnetic unit cell. This has been confirmed experimentally in MnSi by means of Renninger scans in SANS \cite{Adams:PRL2011} and the topological Hall signal \cite{neub09}. We therefore interpret the A-phase in {\cso} as a skyrmion lattice consistent with the study of thin samples \cite{Tokura:2011}, where microscopic proof for the correct winding number is beyond the scope of our study. 

We also confirmed that the temperature and field range of the SANS patterns shown in Fig.\,\ref{Fig3} correspond with the phase diagram in Fig.\,\ref{Fig2}. Typical $T$ and $B$ dependencies of peak intensities are shown in Fig.\,\ref{Fig4} (as peak intensities are shown absolute intensities may not be compared easily in this plot). We find that: (i) the helical order at $B=0$ is characteristic of a second order phase transition at $T_c$ (Fig.\,\ref{Fig4}\,(a)); (ii) the transition from the helical to the conical state is at $B_{c1}$ and the suppression of the conical state at $B_{c2}$ (cf. Fig.\,\ref{Fig4}\,(b)); (iii) in the A-phase the signal of the conical phase vanishes completely in a narrow range (Fig.\,\ref{Fig4}\,(e)). It seems likely that the regime of the coexistence of the conical phase and the A-phase depends on demagnetising fields and thus the sample shape as recently observed in MnSi \cite{Bauer:2012}).

\begin{figure}
\includegraphics[width=0.45\textwidth]{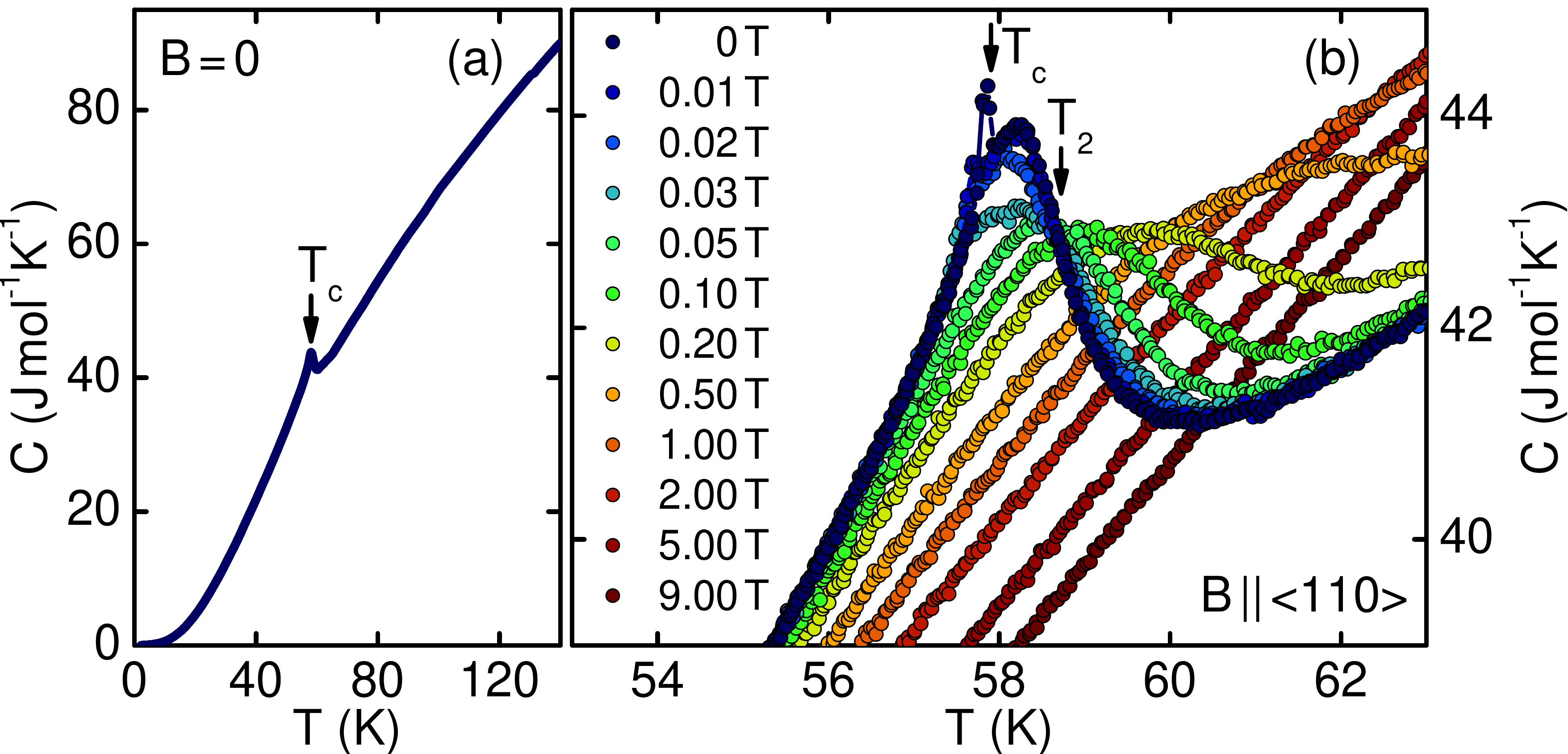}
\caption{(Colour online) Specific heat, $C$, of {\cso}. With increasing temperature the helimagnetic transition is characterised by peak at $T_c$ followed by a hump with a point of inflection at $T_2$. The lack of field dependence at $T_2$ is known as Vollhardt invariance. The $B$ and $T$ dependence of $C$ is characteristic of nearly critical helimagnetic spin fluctuations.}
\label{Fig5}
\end{figure}

We return now to the origin of the enhanced MDR in {\cso}. At the accuracy of our SANS data the transition at $T_c$ is second order (cf. Fig.\,\ref{Fig4}\,(a)). However, the specific heat, $C$, shown in Fig.\,\ref{Fig5}\,(a) and (b), reveals that the transition consists of a narrow peak at $T_c$, characteristic of the latent heat of a first order transition, and a broad hump with a point of inflection at $T_2$. Under applied magnetic fields the peak and hump are suppressed with a shift of entropy towards high temperatures, while the point of inflection at $T_2$ is invariant for $B\lesssim\,B_{\rm c2}$ characteristic of a Vollhardt invariance at $T_2$ \cite{Vollhardt:PRL1997}.  For the case of MnSi it has been shown, that the same behaviour is due to a fluctuation-induced first order transition, where the chiral helimagnetic character of the fluctuations becomes dominant for $T<T_2$. \cite{Stishov:PRB2007,Hamann:PRL2011,Grigoriev:PRB2010,Janoschek:preprint,Bauer:PRB2010}. In turn, this suggests that the enhanced MDR arises from spin currents associated with the helimagnetic character of the spin fluctuations. As the MDR is quantitatively rather small, a full account connecting the magnetic with the dielectric susceptibility poses a challenge for future studies. Interestingly, the enhanced MDR in {\dto} \cite{Saito:PRB2005} quite likely originates in similar spin currents. However, the chiral character of the spin fluctuations thereby must originate in geometric frustration rather than DM interactions. 

The remarkable consistency of the magnetic properties of bulk samples of {\cso} reported here with the binary B20 compounds is a surprise as the unit cell of {\cso} contains 28 instead of 8 atoms. Thus, to the best of our knowledge, bulk samples of {\cso} represent the first example of helimagnetic order in a non-binary B20 compound, a B20 oxide, a B20 compound with a non-ferromagnetic leading order exchange interaction and a B20 insulator. The helimagnetism is thereby consistent with the proposal that thermal Gaussian fluctuations stabilise the skyrmion lattice \cite{mueh09}. In fact, our study clearly contrasts a theoretical proposal \cite{roes06,bute10} in which softened amplitude fluctuations stabilise the skyrmion lattice, since the softening would differ strongly between metals and insulators and with it the extent of the skyrmion lattice phase. Further, the helimagnetism resolves the open questions concerning the magnetisation of {\cso}, offering an explanation for enhancements of the MDR even for non-multiferroic systems such as {\dto}.  Perhaps most importantly, however, as an insulator the skyrmion lattice in {\cso} promises an emergent electrodynamics akin that observed in its binary metallic siblings \cite{joni10,Schulz:NaturePhysics2012}, where electric fields may now be used to manipulate the skyrmions \cite{Seki:2012}.

We wish to thank P. B\"oni, M. Garst, R. Georgii, M. Halder, H. Kolb, S. Mayr, J. Peters, W. Petry and A. Rosch for support and stimulating discussions. Financial support through DFG TRR80, FOR960, the NTH School "Contacs in Nanosystems" and ERC AdG (291079) TOPFIT are gratefully acknowledged. TA, AC, MW, AB and GB acknowledge support through the TUM Graduate School.

%\bibliography{Cu2OSeO3}

\begin{thebibliography}{40}
\bibitem[{Sek()}]{Seki:2012-full}
\bibinfo{note}{Determination of the correct magneto-electric coupling requires
  measurements of the full coupling tensor, which is beyond the scope of our
  study.}
\end{thebibliography}
%\end{document}

\end{document}